\documentclass[journal,10pt]{IEEEtran}
\usepackage{cite}
\usepackage{amsmath,amssymb,amsfonts,stackrel,eqnarray,amsthm}
\usepackage{algorithmic}
\usepackage{graphicx}
\usepackage{textcomp}
\usepackage{xcolor}
\usepackage{multirow}
\usepackage{steinmetz}
\usepackage{amssymb}
\usepackage{tikz}
\usepackage[font=footnotesize]{caption}
\usepackage{subcaption}

\usepackage[T1]{fontenc}

\usepackage{pgfplots}
\pgfplotsset{compat=newest}
\usepackage{cuted}
\newcommand*{\dif}{\mathop{}\!\mathrm{d}}

\DeclareMathOperator{\diag}{diag}

\DeclareMathOperator{\Var}{Var}

\def\BibTeX{{\rm B\kern-.05em{\sc i\kern-.025em b}\kern-.08em
    T\kern-.1667em\lower.7ex\hbox{E}\kern-.125emX}}

\usepackage{siunitx}
\title{Ambient Backscatter Communication in LTE Uplink Sounding Reference Signal}

\author{Jingyi Liao,
    Tianshu Zhang,
    Kalle Ruttik,
    Riku J\"{a}ntti,
    and Dinh-Thuy Phan-Huy
    \thanks {
        J. Liao, K. Ruttik, and R. J\"{a}ntti are with
        Department of Information and Communications Engineering,
        Aalto University,
        02150 Espoo, Finland
        (email: \{firstname.lastname\}@aalto.fi).}
    \thanks{
        Phan-Huy Dinh-Thuy is with
        Networks, Orange Innovation, Ch\^{a}tillon, France
        (email: dinhthuy.phanhuy@orange.com).}
    }
\begin{document}

\maketitle

\begin{abstract}
Ambient Internet of Things (AIoT), recently standardized by the 3rd Generation Partnership Project (3GPP), demands a low-power wide-area communication solution that operates several orders of magnitude below the power requirements of existing 3GPP specifications. Ambient backscatter communication (AmBC) is considered as a competitive potential technique by harvesting energy from the ambient RF signal. This paper considers a symbiotic AmBC into Long Term Evolution (LTE) cellular system uplink. Leveraging by LTE uplink channel estimation ability, AIoT conveys its own message to Base Station (BS) by modulating backscatter path. We explore the detector design, analyze the error performance of the proposed scheme, provide exact expression and its Guassian approximation for the error probability. We corroborate the receiver error performance by Monte Carlo simulation. Analysis of communication range reveals AmBC achieves a reasonable BER of order of magnitude $10^{-2}$ within four times wavelength reading distance. In addition, a AmBC prototype in LTE uplink confirms the its feasibility. The over-the-air experiment results validate theoretical analysis. Hence, the proposed AmBC approach enables AIoT deployment with minimal changes to the LTE system.
\end{abstract}

\section{Introduction}
The Ambient Internet of Things (AIoT) introduces the concept of zero-energy or energy-autonomous devices, driving the development of ultra-low-power communication technologies, which have been investigated by 3GPP \cite{3gppAIoT}. According to this study \cite{3gppAIoT}, Ambient Backscatter Communication (AmBC) is a promising technique for next-generation wireless systems, enabling communication using only energy harvested from far-field radio-frequency (RF) signals present in the environment, such as Digital Video Broadcasting (DVB), Wi-Fi, and cellular transmissions.

The RF signal in cellular communication system carries not only information, but also power. AmBC devices do not generate their own power; instead, they passively reflect and modulate ambient cellular signals. As purely passive components without complex RF front-end circuitry, these devices achieve significant reductions in both cost and energy consumption. 

In Release (Rel.) 18, 3GPP introduced the concept of Ambient Internet of Things (AIoT) as part of 5G-Advanced, targeting low-power, wide-area IoT applications such as inventory tracking, sensing, positioning, and command operations \cite{3gppAIoT}. This zero-energy device offers significantly higher energy efficiency than existing 3GPP Low Power Wide Area (LPWA) solutions, including Narrowband-IoT (NB-IoT)\cite{3gpp_ts_36_300}, LTE Machine Type Communication Category M1 (LTE-MTC, a.k.a. LTE-M) \cite{3gpp_ts_22_368} device, and Reduced Capability (RedCap) \cite{3gpp_ts_38_300} device introduced in Rel. 13 and Rel. 17 respectively. AmBC aligns with the upcoming AIoT initiative in Rel. 19 core topics, scheduled for standardization in 2025 \cite{3gpp_tsg_sa_102}. However, discussions on AIoT capability and detailed specification, including the air inference and corresponding topology, are ongoing in Rel. 20 subsequently. Evolved from 5G enhancement, the sustainable AIoT is excepted to emerge as a critical technology for 6G systems in Rel. 21.

\begin{figure}
    \centering
    \begin{subfigure}[b]{0.55\columnwidth}
    \centering
    \includegraphics[trim={10.5cm 7cm 9.5cm 1.5cm},clip=true,width=\columnwidth]{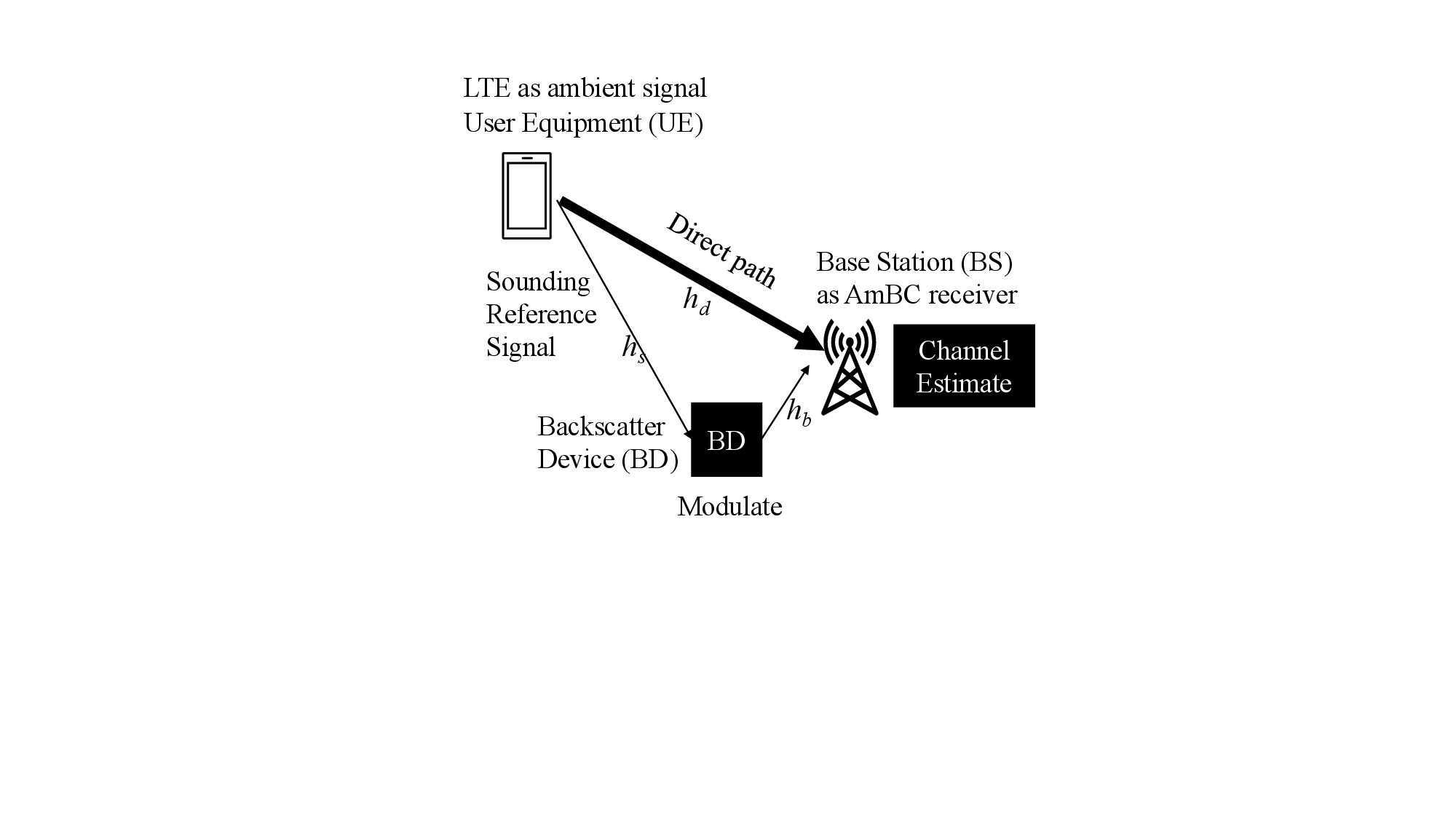}
    \caption{}
    \label{fig:toplevel}
    \end{subfigure}    
    \hfill    
    \begin{subfigure}[b]{0.4\columnwidth}
    \centering
    \includegraphics[width=\columnwidth,trim={1cm 1.5cm, 0, 0},clip]{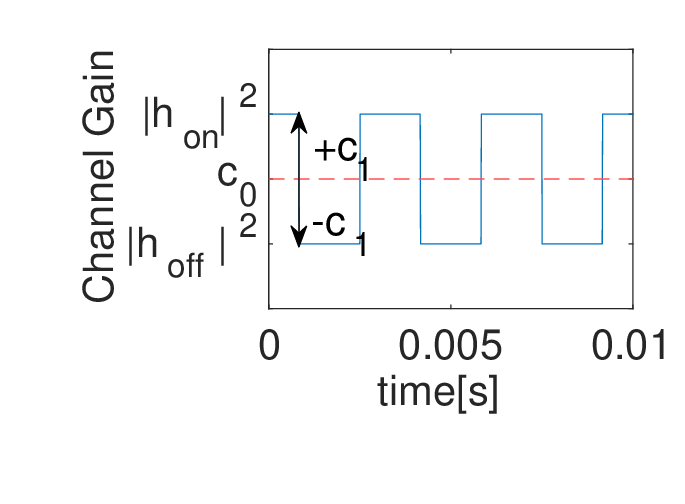}
    \caption{}
    \label{fig:|h[l]|^2}
    \end{subfigure}
    \caption{(a) Illustration of LTE cellular based AmBC system. (b) Channel gain $|h[l]|^2$ is a square-wave.}    
    \vspace{-11pt}
\end{figure}

This paper demonstrates that AmBC can be integrated into LTE network with minimal software update, avoiding significant changes to the existing LTE framework. Our study investigates a configuration similar to symbiotic radio, where a secondary backscatter system shares bandwidth and power with a cellular system. The proposed system aligns with AIoT topology 3, as illustrated in Fig.~\ref{fig:toplevel}. Building upon our previous work \cite{10136397}, which focused on LTE downlink, this study extends the scope to incorporate backscatter communications in both LTE downlink and uplink. Our proposed system leverages periodic reference signals transmitted by the UE for estimating the channel and decoding the backscatter message, enabling the Base Station (BS) to receive backscatter-modulated messages independently of the uplink data traffic. 
We provide a details analysis of receiver detection and coverage performance through both theoretical modeling and numerical simulations.
With a strong emphasis on practical implementation, this study demonstrates the system feasibility using actual UE, marking a significant step towards integrating AmBC systems into existing LTE infrastructure.

\subsection{Related works}
In the 3GPP studies towards Rel 19, there are monostatic backscatter and bistatic backscatter. The monostatic backscatter is similar to Radio Frequency Identification (RFID), whose communication range is shorter than that of bistatic backscatter. In addition, the monostatic backscatter needs separate carrier signal and more complex RF front hardware design \cite{8618337}. To avoid the challenge of communication range and complex transceiver RF front, a bistatic ambient backscatter (topology 3 \cite{3gppAIoT}) is especially attractive technique without independent signal generation (device catalog B \cite{3gppAIoT}).

The first ambient backscatter research using TV broadcast sources as ambient RF signal \cite{10.1145/2486001.2486015}. The Backscatter Device (BD) modifies the incident ambient RF signal by dynamically altering its reflection coefficient. One of the most classical and popular BD is allowed to reflect or absorb RF signal, called On-Off Keying (OOK) scheme \cite{10.1145/2486001.2486015, 10.1145/2619239.2626319}. Existence of BD provides an extra multipath between transmitter and receiver, which modulates the received power on receiver side.

The use of OFDM symbols as ambient RF signals for backscatter communication has been widely explored \cite{8642363, 7551180}, with BD signal extraction based on Received Signal Strength Indicator (RSSI) and Channel State Information (CSI). When relying solely on RSSI, the direct path power is typically several orders of magnitude higher than that of the backscattered path, making direct path interference a significant challenge in AmBC receiver design. To mitigate this issue, several studies have proposed various techniques \cite{8274950, 8103807, 10.1145/2740070.2626312,10.1145/2785956.2787490}. We show special interest in the frequency shift method \cite{194920, 10.1145/2934872.2934894, 10.1145/2934872.2934901, 201552}, shifting to out-of-band. This out-of-band frequency shift scheme requires oscillators, which consume tens to several hundred microwatts. This power consumption is incompatible with the energy-autonomous vision of AIoT, which is constrained to just a few microwatts. In addition to the wide band characteristic of the BD, generating a frequency-shifting signal in the tens of megahertz in the backscatter path could unintentionally cause co-channel interference to other communication systems.

To mitigate out-of-band frequency shift challenges, the BD employs an in-band frequency shift below \(1\,\mathrm{kHz}\), eliminating the need for a precise clock signal in AIoT devices and reducing power consumption to a few microwatts. In-band frequency shift avoids co-channel interference. Also, LTE’s channel estimation and equalization mechanisms inherently track and compensate for in-band frequency shifts. The LTE uplink pilot signal, independent of data traffic, enables continuous channel estimation at a rate exceeding natural Doppler variations\footnote{Typically, natural Doppler is below $100\,\mathrm{Hz}$.}, effectively oversampling the channel. This oversampling creates a gap between the natural Doppler bandwidth and the channel estimator’s bandwidth, which can be exploited to carry BD information. As a result, the BD signal appears as an additional multipath component with an artificial Doppler shift, allowing the AmBC receiver to extract BD information via LTE channel estimators while the LTE BS inherently tracks and mitigates BD signals through equalization, ensuring minimal interference with LTE communications.

By leveraging ambient RF signals, AmBC is seamlessly integrated into cellular networks with minimal tolerable interference \cite{ruttik2018does}. It enables additional IoT communication without disrupting legacy systems and can even enhance channel capacity in existing networks \cite{7948789}. Furthermore, AmBC receivers in LTE do not require additional channel estimation modules, as channel estimation is already embedded in cellular specifications \cite{3gpp_ts_36_211} and will continue to be a fundamental component in future-generation systems.

Our previous work explored AmBC using the LTE downlink pilot signal, specifically the Cell-specific Reference Signal (CRS) \cite{10136397}. However, AmBC performance is constrained by the high path loss between the BS and BD, as LTE downlink Reference Signal Received Power (RSRP) is often insufficient, particularly in indoor environments where BS density is limited \cite{10683260}. To address this limitation, the LTE uplink SRS serves as a complementary solution, as the BS has greater processing power and lower receiver sensitivity than the UE, enhancing detection capabilities. This seamless integration with the cellular system improves connectivity and reliability for AIoT applications. Furthermore, combining downlink and uplink channel estimation can enhance channel diversity and BER performance by leveraging the wideband characteristics of BD. With the increasing adoption of Time Division Duplex (TDD) in 5G and beyond, channel reciprocity necessitates AmBC detection in both downlink and uplink directions \cite{8694831}.
 
\subsection{Contributions}
Existing research have covered on theorems and simulations, and measurements. But to the best of knowledge, however, an investigation on AmBC based on LTE uplink pilot signal is still lacking. This research of LTE uplink pilot signal provides a potential of full duplex uplink / downlink capability, extending the work \cite{10136397}. 

In this paper, we make following contributions:
\begin{enumerate}
    \item Introduce LTE uplink SRS as ambient signal. We utilize the predefined and periodical SRS in LTE uplink. The channel estimates is repurposed to extract AmBC without modification of LTE cellular system.
    \item Analysis of BER performance based on LTE uplink: We conduct a numerical evaluation of the BER performance of the proposed AmBC scheme. This analysis specifically focuses on different approaches to approximate complex distributions.
    \item Analysis bistatic AmBC detection range. We analyze the bistatic backscatter communication range in LTE uplink with fixing BS and UE location.
    \item Over-the-air measurement of the proposed AmBC scheme. We conduct an indoor proof-of-concept measurement using a commercial UE SRS to evaluate BER performance of AmBC.
\end{enumerate}

This paper is organized as follows. Section \ref{sec:System Model} introduces the system model and the problem formulation. Section \ref{sec:SRS} conducts the probability model of received signal and its Gaussian approximation. Section \ref{sec:receiver} designs receivers based on the accuracy and asymptotic probability models. Section \ref{sec:coverage} analyzes AmBC coverage under different uplink central frequency when locations of UE and BS are fixed. Section \ref{sec:Simulation} presents numerical simulations and over-the-air measurements to evaluate the proposed AmBC system. Finally, we conclude the paper in Section \ref{sec:Conclusion}.

\section{System Model \label{sec:System Model}}

\begin{figure}
    \centering  
    \includegraphics[width=0.7\columnwidth]{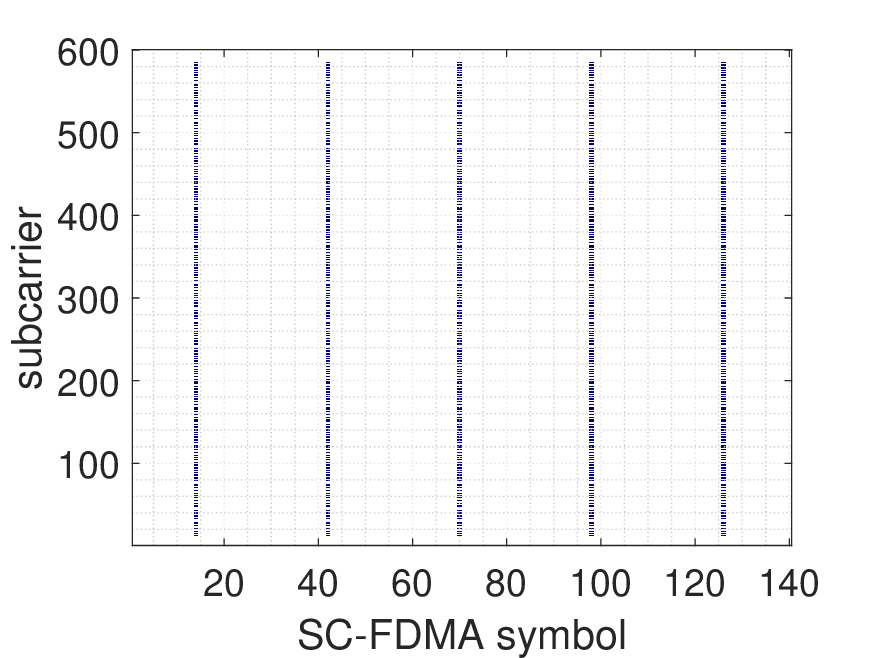}
    \caption{Resource grid of LTE uplink Sounding Reference Signal (SRS).}
    \label{fig:TxRE}
    \vspace{-15pt}
\end{figure}

In Additive White Gaussian Noise (AWGN) channel received signal $s_{r}(t)$ is modeled as 
\begin{eqnarray}
s_{r}(t) = h(t) s_{t}(t) + n(t),
\end{eqnarray}
where $h(t)$ is the channel, $s_{t}(t)$ is transmitted signal and $n(t)$ is AWGN noise with power $P_n=\sigma_n^2$.
The channel contains two parts, a direct path $h_d(t)$ and a scattered path. The scattered path is controlled by BD state $b(t)\in\{-1,1\}$.
The channel from UE to BS is denoted by $h_r(t)$, channel from UE to BD is denoted by $h_s(t)$, and channel from BD to BS is denoted by $h_s(t)$. Channel $h(t)$ represents as
\begin{eqnarray}
h(t) = h_d(t) + h_s(t)h_b(t)[b(t)+1]/2.
\end{eqnarray}

Assume that the state of the backscatter remains constant during a specific Orthogonal Frequency-Division Multiplexing (OFDM) symbol. The received OFDM symbol power is estimated from signal samples \textit{Random Variable} (R.V.) $s_r[i]$, such as
\begin{eqnarray}
\sum_{i=0}^{K-1} |s_{r}[i]|^2 = |h|^2 \sum_{i=0}^{K-1} |s_{t}[i]|^2 + w^2,
\end{eqnarray}
where $i$ is the index of time samples within the OFDM symbol, $K$ is the total number of samples in one OFDM symbol, and the noise $w \sim \mathcal{C}\mathcal{N}(0,\sigma_n^2)$.
The average received power is related to channel gain $|h|^2$, which remains constant during the coherent time correponding to the OFDM symbol duration \cite{10136397}. 
By Parseval's theorem, the received power estimate can equivalently be performed in the frequency domain by averaging over the OFDM subcarriers
\begin{eqnarray} \label{eq: frequency domain}
\sum_{k=0}^{K-1} |S_\text{r}[k]|^2 = |h|^2 \sum_{k=0}^{K-1} |S_\text{t}[k]|^2 + w^2,
\end{eqnarray}
where $S_\text{r}[k]$ and $S_\text{t}[k]$ represent the frequency-domain components of the signal, \textit{i.e.} OFDM symbol subcarries, $k$ is the index in frequency subcarrier.
In Eq. \eqref{eq: frequency domain}, we assume that the BD signal on the scattered path contributes to the first channel tap. The frequency response of this first channel tap is flat, with the justification for this model is detailed in \cite{9923964}. Additionally, the validity of the first channel tap in Eq. \eqref{eq: frequency domain} is corroborated through an alternative LTE uplink channel estimation method described in \cite{5349169}.

Channel gain $|h|^2$ is related to BD states, `on' ($b=1$) and `off' ($b=-1$), for $l$-th OFDM symbol
\begin{eqnarray}
    h[l] = h_d + \frac{b[l]+1}{2} h_s h_b.
\end{eqnarray}
The receiver can distinguish between the two power levels $|h_\text{on}|^2=|h_d + h_s h_b|^2$ and $|h_\text{off}|^2=|h_d|^2$.

\section{SRS channel estimation \label{sec:SRS}}
In cellular systems, a variety of reference signals are transmitted in the uplink to provide CSI. These predefined signals are designed to capture parameters such as multipath fading, scattering, Doppler effects, path loss, and so on. For example, in LTE, the uplink pilot signals include the Sounding Reference Signal (SRS) and the Demodulation Reference Signal (DM-RS) associated with the Physical Uplink Shared Channel (PUSCH). This paper focuses on the LTE uplink SRS due to its key advantage of being independent of uplink traffic, unlike the DM-RS, which is confined to resources allocated for the PUSCH.
Similarly, in New Radio (NR), uplink reference signals include the SRS \cite{3gpp_ts_38_211} (with more configuration options than LTE SRS), the DM-RS, and the Phase Tracking Reference Signal (PT-RS) \cite{3gpp_ts_38_211}. However, we only study LTE case as an example of AmBC integrated in uplink. On the one hand, LTE is more pervasive deployed than NR. On the other hand, AmBC is more suitable working in lower frequency band due to lower path loss. The proposed channel estimation algorithm can be readily extended to other uplink pilot signals in future studies.

The SRS is consisted of a cyclically shifted Zadoff-Chu sequence \cite{5349169}.
It exhibits three notable features.
\begin{enumerate}
    \item \textbf{Constant Unit Amplitude in Single-carrier Frequency-division Multiple Access (SC-FDMA):} Each Resource Element (RE) has a constant unit amplitude.    
    \item \textbf{Consistent Number of REs per SRS:} The SRS OFDM symbol has the same number of REs, $M_\textrm{sc}$.
    \item \textbf{Fixed Transmission Interval:} The SRS is transmitted at a consistent time interval, denoted as $T_\text{SRS}$.
\end{enumerate}
Those three features are critical in the following derivation.

\subsection{Estimation from SRS \label{subsec:one OFDM}}
A RE of $k$-th subcarrier, $l$-th OFDM symbol can be modeled by a noncentral complex Gaussian distribution R.V. $S_\text{r}[k;l]\sim\mathcal{CN}(h[l] S_\text{t}[k;l],\sigma_n^2)$. 
The power of the received element resource $|S_\text{r}[k;l]|^2$ obeys \textit{noncentral chi-square distribution with two degrees of freedom} \cite[Eq. (2.1-140)]{proakis2002communication}, denote as $\chi^{\prime2}_2$.
The $\chi^{\prime2}_2$ are usually discussed with normalized variance Gaussian distribution. For further discussion convenient, we scale R.V. $S_\text{r}[k;l]$ by $\sigma_n/\sqrt{2}$ to the standard normal distribution, 
\begin{equation}
    \frac{\mathfrak{Im}\{S_\text{r}[k;l]\}}{\sigma_n/\sqrt 2} \sim \mathcal{N}
    \left(\frac{\mathfrak{Im}\{h[l]S_\text{t}[k;l]\}}{\sigma_n/\sqrt 2 },1\right),
\end{equation}
\begin{equation}
    \frac{\mathfrak{Re}\{S_\text{r}[k;l]\}}{\sigma_n/\sqrt 2} \sim \mathcal{N}
    \left(\frac{\mathfrak{Re}\{h[l]S_\text{t}[k;l]\}}{\sigma_n/\sqrt 2 },1\right),
\end{equation}
so that 
\begin{equation}
    \frac{|S_\text{r}[k;l]|^2}{\sigma_n^2/2} \sim \chi^{\prime2}_2
    \left(\frac{|h[l]S_\text{t}[k;l]|^2}{\sigma_n^2/2}\right),
\end{equation}
where $\chi^{\prime2}_n(s^2)$ denotes \textit{noncentral chi-square distribution with n degrees of freedom} whose noncentrial parameter is $s^2$.

The amplitude of each SRS RE is unit, \textit{i.e.} $|S_\text{t}[k;l]|^2=1$. Consequently, received LTE uplink signal is simplified as $|S_\text{r}[k;l]h[l]|=|h[l]|$, where the channel gain $h[l]$ remains a constant scalar within the $l$-th OFDM symbol.

Define the R.V. $y[l]$ as the sum of SRS power for the $l$-th OFDM symbol
\begin{equation}
    y[l]=\sum_{k;l\in\text{SRS}} |S_\text{r}[k;l]|^2.
\end{equation}
For a given SRS configuration, the length of SRS signal sequence is always $M_\textrm{sc}$, as specified in the LTE standard \cite[Ch. 5.5.3]{3gpp_ts_36_211}. In other words, the number of SRS RE per OFDM symbol is constant. R.V. $y[l]$, when scaled by noise power $\sigma^2_n/2$, follows noncentral chi-square distribution with $2M_\textrm{sc}$ degrees of freedom whose noncentrial parameter is $2M_\textrm{sc}|h[l]|^2/\sigma_n^2$ \cite{proakis2002communication}
\begin{equation}\label{eq:yChiSquare}
    \frac{y[l]}{\sigma_n^2/2} \sim \chi^{\prime2}_{2M_\textrm{sc}}
    \left(\frac{M_\textrm{sc}|h[l]|^2}{\sigma_n^2/2}\right).
\end{equation}

The R.V. $y[l]$ is \textit{independent and identically distribution (i.i.d.)} for given channel $h_\text{on}$ and $h_\text{off}$. R.V. $y[l]$ distribution related to backscatter state $b$. The \textit{probability density function (p.d.f.)} of $y[l]$ is \cite{proakis2002communication}
\begin{equation}
\begin{split}
    &p\left(\frac{y[l]}{\sigma_n^2/2}\right) =
    \frac{1}{2} \left(\frac{y[l]\sigma_n^2/2}{M_\textrm{sc}|h[l]|^2}\right) ^{\frac{M_\textrm{sc}-1}{2}}\\
    &\exp\left[-\frac{M_\textrm{sc}|h[l]|^2}{\sigma_n^2}-\frac{y[l]}{2}\right]
    I_{M_\textrm{sc}-1} \left(\sqrt{\frac{M_\textrm{sc}|h[l]|^2y[l]}{\sigma_n^2/2}}\right),
\end{split}
\end{equation}
where $I_\alpha(x)$ is the \textit{modified Bessel function of the first kind and order $\alpha$} \cite[Ch. 9.6.1]{abramowitz2012handbook}.

\subsection{Approximation from Noncentrial Chi-square to Gaussian \label{subsec: chi2_gaussian}}
When $M_\textrm{sc}$ is sufficiently large, the R.V. $y[l]$ approaches a Gaussian distribution \cite{7829259}.
For an individual RE, $|S_\text{r}[k;l]|^2$ follows noncentrial chi-square distribution
\begin{equation}
    \frac{|S_\text{r}[k;l]|^2}{\sigma_n^2/2} \sim \chi'^2_2(|h[l]|^2),
\end{equation}
whose mean and variance are
\begin{equation}
    \mathbb{E}\left(|S_\text{r}[k;l]|^2\right) = \sigma_n^2 + |h[l]|^2,
\end{equation}
\begin{equation}
    \Var\left(|S_\text{r}[k;l]|^2\right) = \sigma_n^4 + 2\sigma_n^2|h[l]|^2,
\end{equation}
According to the Central Limit Theorem (CLT), the \textit{Cumulative Distribution Function (c.d.f.)} of $y[l]$ can be approximated as \cite{blitzstein2019introduction} 
\begin{equation} \label{eq:CLT}
    \lim_{M_\textrm{sc}\rightarrow\infty} \text{Pr} \left(\frac{y[l]-\mathbb{E}(y[l])}{\sqrt{\Var(y[l])}}\leq \xi\right) = \Phi(\xi),
\end{equation}
where $\Phi(\xi)$ denotes the \textit{c.d.f}. of the standard normal distribution. Thus, $y[l]$ is approximated by Gaussian distribution
\begin{equation} \label{eq:yGaussian}
    y[l] \rightarrow \mathcal{N}\left(M_\textrm{sc}(\sigma_n^2+ |h[l]|^2),M_\textrm{sc}(\sigma_n^4 +2\sigma_n^2 |h[l]|^2)\right),
\end{equation}
where mean and variance of $y[l]$ are
\begin{equation}
    \mathbb{E}(y[l]) = M_\textrm{sc}(\sigma_n^2+ |h[l]|^2),
    \label{eq:GaussianMean}
\end{equation}
\begin{equation}
    \Var(y[l])
    = M_\textrm{sc} (\sigma_n^4 +2\sigma_n^2 |h[l]|^2).
    \label{eq:GaussianVar}    
\end{equation}
The mean $\mu_y[l]$ and variance $\sigma^2_y[l]$ both depend on a backscatter states `on' and `off'.

\section{Receiver decision \label{sec:receiver}}
The BD generates symbols consisting of the $N$ samples symbols, denoted as $\boldsymbol{s_m}$, based on a square-wave pattern, where $N$ is an even number. Each symbol contains an equal number of on-state ($\boldsymbol{s_m}[i]=1$) and off-state ($\boldsymbol{s_m}[i]=-1$) samples. The BD transmits digital information using two distinct waveforms $\boldsymbol{s_m}(t), m\in\{0, 1\}$ within the symbol interval $T_s$. Sampling is performed at intervals corresponding to the SRS time interval, $T_\text{SRS}$. Discrete sequence of signal $\boldsymbol{s_m}$ and received symbol $\boldsymbol{y}$ are column vectors with $N$ real elements.

In this section, we derive the optimal receiver based on detection probability. We then examine several alternative receivers that approximate the optimal design, including square root receivers, correlation receivers, and power receivers. Additionally, we provide a detailed analysis of the detection error probabilities of correlation receivers and their BER Gaussian interpretation.

\subsection{Noncentral Chi-square Distribution}
\subsubsection{Accuracy Distribution}
The received signal $\boldsymbol{y}$ becomes a N-element R.V. vector. The joint \textit{p.d.f.} of $2\boldsymbol{y}/\sigma^2_n$ is
\begin{equation}
    p\left(\frac{\boldsymbol{y}}{\sigma_n^2/2}\right) = \prod_{i=0}^{N-1} p\left(\frac{y[i]}{\sigma_n^2/2}\right)
\end{equation}
The logarithm likelihood \textit{p.d.f.} of $2\boldsymbol{y}/\sigma_n^2$ is
\begin{align}
    &\ln p\left(\frac{\boldsymbol{y}}{\sigma_n^2/2} | \boldsymbol{s_m}\right)
    = \sum_{i=0}^{N-1} \ln p\left(\frac{y[i]}{\sigma_n^2/2} | \boldsymbol{s_m}\right) =\\
    &N\ln \frac{1}{2}
    + \frac{M_\textrm{sc}-1}{2} \sum_{i=0}^{N-1} \ln\left(\frac{y[i]\sigma_n^2/2}{M_\textrm{sc}|h[i]|^2}\right)
    -\frac{M_\textrm{sc}}{\sigma_n^2}\sum_{i=0}^{N-1}|h[i]|^2 \nonumber\\
    &-\frac{1}{2}\sum_{i=0}^{N-1} y[i] +
    \sum_{i=0}^{N-1} \ln I_{M_\textrm{sc}-1} \left(\sqrt{\frac{M_\textrm{sc}|h[i]|^2y[i]}{\sigma_n^2/2}}\right).
\end{align}
The optimum receiver of a binary-modulated signal operates based on the maximum-a-posteriori (MAP) principle. By applying the maximum likelihood criterion, the optimal decision rule for the receiver is established as
\begin{eqnarray}
\mathrm{Pr}\{\boldsymbol{s_0}|\boldsymbol{y}\}
\mathrel{\substack{\mathcal{H}_1\\ \lessgtr \\ \mathcal{H}_0}}
\mathrm{Pr}\{\boldsymbol{s_1}|\boldsymbol{y}\}
,
\label{eq: MAP}
\end{eqnarray}
while the $\mathcal{H}_0$ is symbol 0 is transmitted and $\mathcal{H}_1$ is symbol 1 is transmitted. According to Bayes' theorem, the Eq. \eqref{eq: MAP} is 
\begin{eqnarray}
\frac{p(\boldsymbol{y}|\boldsymbol{s_0})\mathrm{Pr}\{\boldsymbol{s_0}\}}{p(\boldsymbol{y})}
\mathrel{\substack{\mathcal{H}_1\\ \lessgtr \\ \mathcal{H}_0}}
\frac{p(\boldsymbol{y}|\boldsymbol{s_1})\mathrm{Pr}\{\boldsymbol{s_1}\}}{p(\boldsymbol{y})}
.
\end{eqnarray}
The logarithm likelihood functions give a hypothesis
\begin{equation}
    \ln p\left(\frac{\boldsymbol{y}}{\sigma_n^2/2} | \boldsymbol{s_0}\right)
    \mathrel{\substack{\mathcal{H}_1\\ \lessgtr \\ \mathcal{H}_0}}    
    \ln p\left(\frac{\boldsymbol{y}}{\sigma_n^2/2} | \boldsymbol{s_1}\right),
\end{equation}
is simplified as
\begin{equation}
\label{eq:BesselMAP}
\begin{split}    
    &\sum_{i=0}^{N-1} \ln I_{M_\textrm{sc}-1} \left( \left| h_d+ h_sh_b \frac{1+s_0[i]}{2}\right| \sqrt{\frac{M_\textrm{sc} y[i]}{\sigma_n^2/2}}\right)\\
    &\mathrel{\substack{\mathcal{H}_1\\ \lessgtr \\ \mathcal{H}_0}} \\
    &\sum_{i=0}^{N-1} \ln I_{M_\textrm{sc}-1} \left( \left| h_d+ h_sh_b \frac{1+s_1[i]}{2}\right| \sqrt{\frac{M_\textrm{sc}y[i]}{\sigma_n^2/2}}\right).
\end{split}
\end{equation}
Eq. \eqref{eq:BesselMAP} represents the optimal receiver involving a modified Bessel function of first kind. However, the order of Bessel function is exceedingly high. As a result, the numerical simulations of that Bessel function run suffer from number precision limitations, rendering the optimal receiver in Eq. \eqref{eq:BesselMAP} is impractical in engineering application.
\subsubsection{Approximation of Bessel Function}
The integral formulas for the modified Bessel functions $I_\alpha(x)$ for $\mathfrak{Re}(x)>0$ is \cite[Eq. (4)]{gray1895treatise}
\begin{eqnarray}
\begin{aligned}
    I_\alpha(x)=&\frac{1}{\pi} \int_0^\pi e^{x\cos \theta}\cos(\alpha\theta) \dif \theta\\
    &- \frac{\sin(\alpha\pi)}{\pi} \int_0^\pi e^{-x\cosh t-\alpha t} \dif t,
\end{aligned}
\end{eqnarray}
where $\alpha=M_\textrm{sc}-1$ is integer so $\sin(\alpha\pi)=0$. Then the integration becomes
\begin{equation}
    I_\alpha(x)=\frac{1}{\pi} \int_0^\pi e^{x\cos \theta}\cos(\alpha\theta) \dif \theta.
\end{equation}
The integrand $e^{x\cos \theta}\cos(\alpha\theta)$ is too complex to integral.
\textbf{For low SNR case, $x\to0$.} The Taylor series of integrand $e^{x\cos \theta}\cos(\alpha\theta)$ expands with $\theta=0$ of first order is $e^x$.
We replace integrand using the its Taylor series first term $e^x$.
\begin{equation}
    I_\alpha(x) \approx \frac{1}{\pi} \int_0^\pi e^x \dif \theta = e^x.
\end{equation}
Put this approximation in Eq. \eqref{eq:BesselMAP},
\begin{equation}
\begin{split}    
    &\sum_{i=0}^{N-1}  \left| h_d+ h_sh_b \frac{1+s_0[i]}{2}\right| \sqrt{\frac{M_\textrm{sc} y[i]}{\sigma_n^2/2}} \\
    &\mathrel{\substack{\mathcal{H}_1\\ \lessgtr \\ \mathcal{H}_0}} \\
    &\sum_{i=0}^{N-1} \left| h_d+ h_sh_b \frac{1+s_1[i]}{2}\right| \sqrt{\frac{M_\textrm{sc}y[i]}{\sigma_n^2/2}}.
\end{split}
\end{equation}
The $\Delta h$ denotes $|h_d + h_s h_b|-|h_d|$. Subsequently, the approximation of Eq. \eqref{eq:BesselMAP} becomes
\begin{equation}
\label{eq:SqrtMAP}
    \sum_{i=0}^{N-1} s_0[i] \Delta h \sqrt{\frac{y[i]}{\sigma_n^2/2}}
    \mathrel{\substack{\mathcal{H}_1\\ \lessgtr \\ \mathcal{H}_0}} 
    \sum_{i=0}^{N-1} s_1[i] \Delta h \sqrt{\frac{y[i]}{\sigma_n^2/2}}.
\end{equation}
\textbf{For high SNR case, $x\to\infty$.} The asymptotic series expansions of $I_\alpha(x)$ is \cite{BesselI_inf}
\begin{equation}
    I_\alpha(x) \propto \frac{e^x}{\sqrt{2\pi x}}\left[1+\mathcal{O}\left(\frac{1}{x}\right)\right].
\end{equation}
The residual terms are ignored, then the logarithm of modified Bessel function becomes
\begin{equation}
    \ln I_\alpha(x)\propto x-\frac{1}{2}\ln{x}.
\end{equation}
Finally, the approximation of Eq. \eqref{eq:BesselMAP} draws the same conclusion
\begin{equation}
    \sum_{i=0}^{N-1} s_0[i] \Delta h \sqrt{\frac{y[i]}{\sigma_n^2/2}}
    \mathrel{\substack{\mathcal{H}_1\\ \lessgtr \\ \mathcal{H}_0}} 
    \sum_{i=0}^{N-1} s_1[i] \Delta h \sqrt{\frac{y[i]}{\sigma_n^2/2}}.
\end{equation}
We get the same approximation as Eq. \eqref{eq:SqrtMAP} from both low and high SNR circumstances.

Compared to Eq. \eqref{eq:BesselMAP}, the square root in Eq. \eqref{eq:SqrtMAP} is more practical but remains challenging to implement in hardware.
For further simplify, a series representation of the square root is
\begin{align}
    \sqrt{x}
    =& \sum_{n=0}^\infty {\frac{1}{2} \choose n} (x-1)^n\\
    =& 1 + \frac{x-1}{2}-\frac{(x-1)^2}{8} + \mathcal{O}(x^3),
    \label{eq:Taylor}
\end{align}
when $x\in \mathbb{R}^+$. Each term in this Taylor series corresponds to a specific moment of the R.V. For example, the first order term of the Taylor series transforms the receiver into a correlator receiver. Including the second order term of Taylor series incorporates the second moment of $y$, leading to a power receiver. The second moment receiver aligns with the Gaussian optimized receiver (see Sec. \ref{subsec:GaussianRx}), when the noncentral chi-square R.V. $y[l]$ asymptotes to a Gaussian distribution.

The simplest receiver is the correlator receiver, derived from the first order of Taylor series Eq. \eqref{eq:Taylor}. In this case, Eq. \eqref{eq:SqrtMAP} simplifies to
\begin{equation}
\label{eq:corrMAP}
    \sum_{i=0}^{N-1} s_0[i] \Delta h \frac{y[i]}{\sigma_n^2/2}
    \mathrel{\substack{\mathcal{H}_1\\ \lessgtr \\ \mathcal{H}_0}} 
    \sum_{i=0}^{N-1} s_1[i] \Delta h \frac{y[i]}{\sigma_n^2/2}.
\end{equation}
This correlator detector is perfected because it is low-cost in hardware application. Its theorem performance is discussed in Sec. \ref{subsec: BER analysis}.

The BPSK represents the simplest and special case where $\boldsymbol{s_0}=-\boldsymbol{s_1}$. Under this condition, the square root detector Eq. \eqref{eq:SqrtMAP} and correlation detector Eq. \eqref{eq:corrMAP} are simplified accordingly as
\begin{equation}
    \sum_{i=0}^{N-1} s_0[i] \Delta h \sqrt{\frac{y[i]}{\sigma_n^2/2}}
    \mathrel{\substack{\mathcal{H}_1\\ \lessgtr \\ \mathcal{H}_0}} 0,
    \label{eq: rxSqrt}
\end{equation}
and
\begin{equation}
    \sum_{i=0}^{N-1} s_0[i] \Delta h \frac{y[i]}{\sigma_n^2/2}
    \mathrel{\substack{\mathcal{H}_1\\ \lessgtr \\ \mathcal{H}_0}} 0.
    \label{eq: rxCorr}
\end{equation}
Indeed, the there are some differences between the square root detector and the correlation detector in their analytical expressions. We compare the performance of two detectors through numerical simulations in Sec. \ref{sec:Simulation}.

\subsection{Asymptotic Gaussian Distribution\label{subsec:GaussianRx}}

In previous section, $y$ follows noncentrial chi-square distribution. In this section, we will derive a receiver based on the approximation that $y$ follows Gaussian distribution.
The symbol denotions are the same as previous section. The received R.V. vector $\boldsymbol{y}$ approximation relies on transmitted symbol $\boldsymbol{s_m}$ (i.e., $\boldsymbol{s_0}$ or $\boldsymbol{s_1}$)
\begin{equation}
    \boldsymbol{y} \rightarrow \mathcal{N} \left( \boldsymbol{\mu_y}, \boldsymbol{C} \right),
\end{equation}
where column vector $\boldsymbol{\mu_y}$ is mean
\begin{equation}
\boldsymbol{\mu_y} = M_\textrm{sc}\left(\sigma_n^2+\left|h_d + \frac{\boldsymbol{s_m}+1}{2} h_s h_b\right|^2\right),
\end{equation}
and square diagonal matrix $\boldsymbol{C}$ is covariance matrix
\begin{equation}
    \boldsymbol{C}=\diag\left\{M_\textrm{sc}\left[\sigma_n^4 +2\sigma_n^2 \left|h_d + \frac{\boldsymbol{s_m}+1}{2} h_s h_b\right|^2\right]\right\}. 
\end{equation}
Above, the $\boldsymbol{y}$, $\boldsymbol{s_m}$, $\boldsymbol{\mu_y}$ are column vectors with $N$ real elements. The $\boldsymbol{C}$ is a real $N \times N$ elements matrix.

The joint \textit{p.d.f.} of random vector $\boldsymbol{y}$ is
\begin{align}
    &p(\boldsymbol{y} | \boldsymbol{s_m})\nonumber\\
    =& \frac{1}{\sqrt{(2\pi)^{N}\det(\boldsymbol{C})}}\exp\left[-\frac{(\boldsymbol{y}-\boldsymbol{\mu_y})^T\boldsymbol{C}^{-1}(\boldsymbol{y}-\boldsymbol{\mu_y})}{2}\right]\\
    =& \left(\frac{2\pi}{M_\textrm{sc}}\right)^{-\frac{N}{2}}
    \prod_{i=0}^{N-1} \left(\sigma_n^4 +2\sigma_n^2 \left|h_d + \frac{s_m[i]+1}{2} h_s h_b\right|^2\right)
    ^{-\frac{1}{2}} \nonumber\\
    &\exp\left[-\frac{1}{2}\sum_{i=0}^{N-1}  \frac{\left(y[i] - \sigma_n^2 - \left|h_d + \frac{s_m[i]+1}{2} h_s h_b\right|^2\right)^2}{M_\textrm{sc}(\sigma_n^4 +2\sigma_n^2 \left|h_d + \frac{s_m[i]+1}{2} h_s h_b\right|^2)} \right]
    .
\label{eq: y_PDF}
\end{align}
Logarithm \textit{p.d.f.} of random vector $\boldsymbol{y}$ is
\begin{align}
    &\ln p(\boldsymbol{y} | \boldsymbol{s_m})={-\frac{N}{2}} \ln \left(\frac{2\pi}{M_\textrm{sc}}\right) \nonumber\\
    &-\frac{1}{2}
    \sum_{i=0}^{N-1} \left(\sigma_n^4 +2\sigma_n^2 \left|h_d + \frac{s_m[i]+1}{2} h_s h_b\right|^2\right)
    \nonumber\\
    &-\frac{1}{2}\sum_{i=0}^{N-1}  \frac{\left(y[i] - \sigma_n^2 - \left|h_d + \frac{s_m[i]+1}{2} h_s h_b\right|^2\right)^2}{M_\textrm{sc}(\sigma_n^4 +2\sigma_n^2 \left|h_d + \frac{s_m[i]+1}{2} h_s h_b\right|^2)}
    .
\label{eq: y_logPDF}
\end{align}
Similarly, the MAP criteria is simplified as
\begin{equation}
\label{eq:GaussianMAP}
\begin{split}    
    &-\frac{1}{2}\sum_{i=0}^{N-1} \frac{y[i]^2}{\sigma_n^2 +2 \left|h_d + \frac{s_0[i]+1}{2} h_s h_b\right|^2}\\
    &+\sum_{i=0}^{N-1} y[i]\frac{\sigma_n^2 + \left|h_d + \frac{s_0[i]+1}{2} h_s h_b\right|^2}{\sigma_n^2 +2 \left|h_d + \frac{s_0[i]+1}{2} h_s h_b\right|^2}\\
    &\mathrel{\substack{\mathcal{H}_1\\ \lessgtr \\ \mathcal{H}_0}} \\
    &-\frac{1}{2}\sum_{i=0}^{N-1} \frac{y[i]^2}{\sigma_n^2 +2 \left|h_d + \frac{s_1[i]+1}{2} h_s h_b\right|^2}\\
    &+\sum_{i=0}^{N-1} y[i]\frac{\sigma_n^2 + \left|h_d + \frac{s_1[i]+1}{2} h_s h_b\right|^2}{\sigma_n^2 +2 \left|h_d + \frac{s_1[i]+1}{2} h_s h_b\right|^2}.
\end{split}
\end{equation}

\begin{figure*}
\begin{equation}
    \sum_{i=0}^{N-1} \frac{\left(\left|h_d + \frac{s_0[i]+1}{2} h_s h_b\right|^2 - \left|h_d + \frac{s_1[i]+1}{2} h_s h_b\right|^2\right)(y[i]^2-\sigma_n^2y[i])}{\left(\sigma_n^2+\left|h_d + \frac{s_0[i]+1}{2} h_s h_b\right|^2\right)\left(\sigma_n^2+\left|h_d + \frac{s_1[i]+1}{2} h_s h_b\right|^2\right)}
    \mathrel{\substack{\mathcal{H}_1\\ \lessgtr \\ \mathcal{H}_0}} 0
\end{equation}
\hrule
\end{figure*}

When $\boldsymbol{y}$ is approximated by a Gaussian distribution, this power receiver is optimal. For BPSK symbol, the receiver represents as
\begin{equation}
    \sum_{i=0}^{N-1} s_0[i] \Delta h \left(\frac{y[i]}{\sigma_n^2/2}-1\right)^2
    \mathrel{\substack{\mathcal{H}_1\\ \lessgtr \\ \mathcal{H}_0}} 0.
    \label{eq: square}
\end{equation}

\subsection{BER analysis \label{subsec: BER analysis}}
Thus far, we have derived three approximate receivers: the square root detector Eq. \eqref{eq: rxSqrt}, correlation detector Eq. \eqref{eq: rxCorr}, and power detector Eq. \eqref{eq: square}.  Each of these receivers is approximated from the exact noncentral chi-square distribution and its optimal receiver Eq. \eqref{eq:BesselMAP}. In this subsection, we focus specifically on analyzing the detection error probabilities of correlation receivers.

Based on correlation detector Eq. \eqref{eq:corrMAP}, only the samples where $s_0[i] \neq s_1[i]$ eliminate uncertainty and contribute to the detector. BPSK is discussed for convenience, as each symbol's samples are distinct. However, a limitation of BPSK is its sensitivity to sign of $\Delta h$. In a slow fading channel, the sign of $\Delta h$ fluctuates over time. To investigate the BER, $P_e$, we assume $\Delta h$ is a positive scalar. As shown in Eq. \eqref{eq:yChiSquare}, the distribution parameters of the R.V. $y[l]$ depend on the BD states. We classify $y[l]$ into two groups: $s_0[l]=1$ and $s_0[l]=-1$. Then Eq. \eqref{eq: rxCorr} is
\begin{equation}
    -\sum_{\substack{i=0,\\s_0[i]=-1}}^{N-1} \frac{y[i]}{\sigma_n^2/2}
    +\sum_{\substack{i=0,\\s_0[i]=1}}^{N-1} \frac{y[i]}{\sigma_n^2/2}
    \mathrel{\substack{\mathcal{H}_1\\ \lessgtr \\ \mathcal{H}_0}}
    0,
\end{equation}
eliminating $\Delta h$ for both sides of inequality.
If the sign remains the same positive or negative in symbol duration, the sign of $\Delta h$ is not critical for the final $P_e$. Because a similar \textit{p.d.f.} can be conducted by flipping the hypothesis test, ultimately leading to the same $P_e$. 

Defined normalized $y_-$ and normalized $y_+$ as
\begin{equation}
    y_-=\sum_{\substack{i=0,\\s_0[i]=-1}}^{N-1} \frac{y[i]}{\sigma_n^2/2},
\end{equation}
\begin{equation}
    y_+=\sum_{\substack{i=0,\\s_0[i]=1}}^{N-1} \frac{y[i]}{\sigma_n^2/2}.
\end{equation}
The sum of noncentral chi-square distributions remains noncentral chi-square distribution \cite{blitzstein2019introduction}. Therefore, based on Eq. \eqref{eq:yChiSquare}, $y_\pm$ are also a noncentral chi-square distribution
\begin{equation}
    y_- | s_0 \sim \chi_{M_\mathrm{sc}N}^{\prime2}\left(N M_\mathrm{sc}\frac{\left|h_d\right|^2}{\sigma_n^2}\right),
\end{equation}
\begin{equation}
    y_+ | s_0 \sim \chi_{M_\mathrm{sc}N}^{\prime2}\left(N M_\mathrm{sc}\frac{\left|h_d + h_s h_b\right|^2}{\sigma_n^2}\right),
\end{equation}
and
\begin{equation}
    y_- | s_1 \sim \chi_{M_\mathrm{sc}N}^{\prime2}\left(N M_\mathrm{sc}\frac{\left|h_d + h_s h_b\right|^2}{\sigma_n^2}\right),
\end{equation}
\begin{equation}
    y_+ | s_1 \sim \chi_{M_\mathrm{sc}N}^{\prime2}\left(N M_\mathrm{sc}\frac{\left|h_d\right|^2}{\sigma_n^2}\right).
\end{equation}

The exact density of the difference of two linear combinations of independent noncentral chi-square R.V.s
\begin{equation}
 -y_- + y_+
    \mathrel{\substack{\mathcal{H}_1\\ \lessgtr \\ \mathcal{H}_0}}
    0
\end{equation}
is expressed in terms of Whittaker's function and provided in closed form \cite{provost1996exact}.
However, the distribution function in \cite[THEOREM 3.1]{provost1996exact} is an infinite series and does not converge as quickly as the doubly noncentral F distribution. In this paper, we recommend using the noncentral F distribution for a more accurate \textit{c.d.f} approximation. This approach is the similar to the one outline in the textbook \cite{kay2009fundamentals}, where we convert the difference of two independent noncentral chi-square R.V. into the ratio of two independent noncentral chi-square R.V.
\begin{equation}
    \xi=\frac{y_+}{y_-}
    \mathrel{\substack{\mathcal{H}_1\\ \lessgtr \\ \mathcal{H}_0}}
    1,
\end{equation}
known as doubly noncentral F distribution \cite{b2955244-f463-3d6e-a420-cdd1c88c4dad}
\begin{equation}
    \xi | s_0 \sim F^{\prime\prime}_{M_\mathrm{sc}N,M_\mathrm{sc}N} \left(
    M_\mathrm{sc}N\frac{\left|h_d + h_s h_b\right|^2}{\sigma_n^2},
    M_\mathrm{sc}N\frac{\left|h_d\right|^2}{\sigma_n^2}
    \right),
\end{equation}
and
\begin{equation}
    \xi | s_1 \sim F^{\prime\prime}_{M_\mathrm{sc}N,M_\mathrm{sc}N} \left(
    M_\mathrm{sc}N\frac{\left|h_d\right|^2}{\sigma_n^2},
    M_\mathrm{sc}N\frac{\left|h_d + h_s h_b\right|^2}{\sigma_n^2}
    \right).
\end{equation}

\begin{figure}
    \centering  
    \includegraphics[width=0.56\columnwidth]{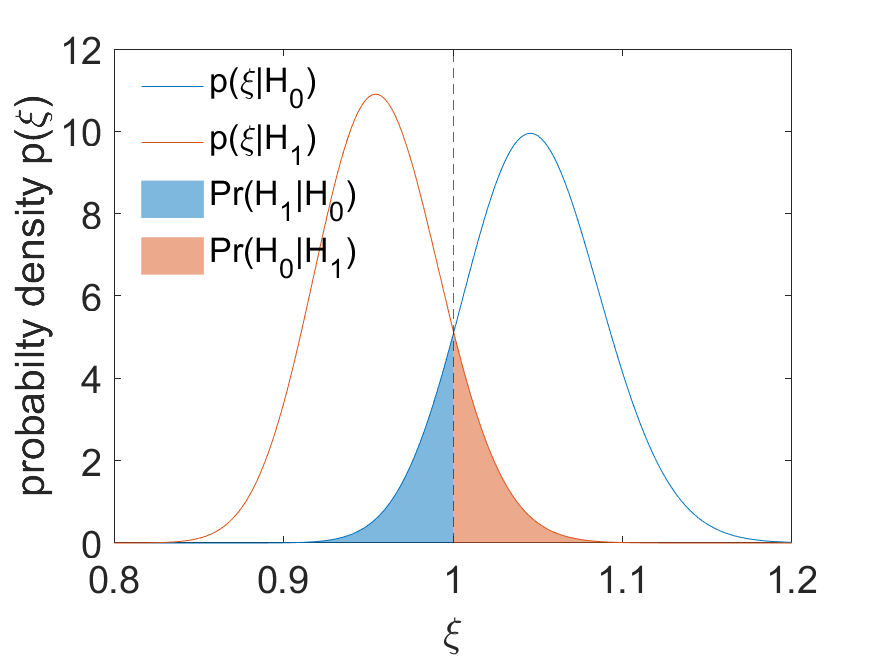}
    \caption{Detection performance of $\xi$ on doubly noncentral F Distribution when LTE SNR is 5 dB.}
    \label{fig:pdfDoublyNoncentralFDistribution}
    \vspace{-15pt}
\end{figure}

The \textit{c.d.f.} of the doubly noncentral F distribution is given in textbook \cite[Eq. (30.51)]{johnson1995continuous} as Eq. \eqref{eq: pdfNoncentralF_0} and Eq. \eqref{eq: pdfNoncentralF_1},
\begin{figure*}
\begin{equation}
\label{eq: pdfNoncentralF_0}
\begin{split}
    \mathrm{Pr}(\xi \leq x|s_0) = 
    &\exp \left[ -\frac{M_\mathrm{sc}N (\left|h_d\right|^2+\left|h_d + h_s h_b\right|^2) }{4\sigma_n^2} \right]\\
    &\sum_{j=0}^\infty \sum_{k=0}^\infty
    \frac{\left(\frac{M_\mathrm{sc}N\left|h_d + h_s h_b\right|^2}{4\sigma_n^2}\right)^j \left(\frac{M_\mathrm{sc}N\left|h_d\right|^2}{4\sigma_n^2}\right)^k}{j!k!}
    \bar{B}_\frac{x}{1+x}
    \left[\frac{M_\mathrm{sc}N}{2}+j,\frac{M_\mathrm{sc}N}{2}+k\right]
\end{split}
\end{equation}
\begin{equation}
\label{eq: pdfNoncentralF_1}
\begin{split}
    \mathrm{Pr}(\xi \leq x|s_1) = 
    &\exp \left[ -\frac{M_\mathrm{sc}N (\left|h_d\right|^2+\left|h_d + h_s h_b\right|^2) }{4\sigma_n^2} \right]\\
    &\sum_{j=0}^\infty \sum_{k=0}^\infty
    \frac{ \left(\frac{M_\mathrm{sc}N\left|h_d\right|^2}{4\sigma_n^2}\right)^j \left(\frac{M_\mathrm{sc}N\left|h_d + h_s h_b\right|^2}{4\sigma_n^2}\right)^k}{j!k!}
    \bar{B}_\frac{x}{1+x}
    \left[\frac{M_\mathrm{sc}N}{2}+j,\frac{M_\mathrm{sc}N}{2}+k\right]
\end{split}
\end{equation}
\hrule
\end{figure*}
where $\bar{B}_x(a,b)$ denotes the regularized Beta function, as defined in \cite[Ch. 26.5.1]{abramowitz2012handbook}.
The detection threshold is set to 1, and $M_\mathrm{sc}N/2$ is an integer in practice. The BER is composed of two types of error
\begin{equation}
    P_e = \mathrm{Pr}(s_0)\mathrm{Pr}(\xi \leq 1|s_0) + \mathrm{Pr}(s_1)[1-\mathrm{Pr}(\xi \leq 1|s_1)],
\end{equation}
as Fig. \ref{fig:pdfDoublyNoncentralFDistribution} shown. For BPSK symbol, the BER is simplified to Eq. \eqref{eq: P_e}.
\begin{figure*}
\begin{equation}
    \label{eq: P_e}
    \begin{split}
        P_e= &\frac{1}{2} +
        \frac{1}{2}
        \exp \left[ -\frac{M_\mathrm{sc}N (\left|h_d\right|^2+\left|h_d + h_s h_b\right|^2) }{4\sigma_n^2} \right] +\\
    &\sum_{j=0}^\infty \sum_{k=0}^\infty
    \left(\frac{M_\mathrm{sc}N}{4\sigma_n^2}\right)^{j+k}
    \frac{\left|h_d + h_s h_b\right|^{2j} \left|h_d\right|^{2k}
    -\left|h_d\right|^{2j} \left|h_d + h_s h_b\right|^{2k}}{j!k!}
    \bar{B}_\frac{x}{1+x}
    \left[\frac{M_\mathrm{sc}N}{2}+j,\frac{M_\mathrm{sc}N}{2}+k\right] .
    \end{split}
\end{equation}
\hrule
\end{figure*}
The BER performance of the FSK coherent detector is comparable to that of BPSK. Due to space constraints, a detailed discussion of FSK is omitted.

\subsection{Connection to Gaussian Distribution}
The Eq. \eqref{eq: P_e} provides the exact error probability but its closed form is mathematically complex. An approximate error probability can be derived by using the asymptotic Gaussian distribution.

As discussed in Sec. \ref{subsec: chi2_gaussian}, the number of SRS subcarrier, $M_\textrm{sc}$, is sufficiently  large to approximate the R.V. $y[l]$ using Gaussian approximation. In textbook \cite[Eq. (5.2-4)]{proakis2002communication}, the classical BER of BPSK in AWGN is given by
\begin{equation}
    \Tilde{P_e} = Q\left(\sqrt{\frac{2\mathcal{E}_b}{N_0}}\right),
\end{equation}
where $Q(\cdot)$ is the tail distribution function of the standard normal distribution. The $\mathcal{E}_b$ and $N_0$ refer to symbol power and noise power density. The textbook \cite[Eq. (5.2-4)]{proakis2002communication} used the same correlator as Sec. \ref{subsec: BER analysis}, assuming AWGN conditions.
In BD signal the $\mathcal{E}_b$ and $N_0$ are
\begin{equation}
    \mathcal{E}_b \approx N M_\textrm{sc}^2 \frac{(|h_d + h_s h_b|^2-|h_d|^2)^2}{4},
\end{equation}
and 
\begin{equation}
    \frac{1}{2}N_0 \approx M_\textrm{sc}\sigma_n^2\left(\sigma_n^2+|h_d + h_s h_b|^2+|h_d|^2\right).
\end{equation}
The SNR per bit $\gamma_b$ is derived by them as
\begin{equation}
    \label{eq: SNR_gamma_b}
    \gamma_b=\frac{\mathcal{E}_b}{N_0}=\frac{N M_\textrm{sc} (|h_d + h_s h_b|^2-|h_d|^2)^2}{8\sigma_n^2\left(\sigma_n^2+|h_d + h_s h_b|^2+|h_d|^2\right)}.
\end{equation}

The error probability of AmBC asymptotes to
\begin{equation}
    \Tilde{P_e} = Q\left(\sqrt{\frac{N M_\textrm{sc} (|h_d + h_s h_b|^2-|h_d|^2)^2}{4\sigma_n^2\left(\sigma_n^2+|h_d + h_s h_b|^2+|h_d|^2\right)}}\right).
    \label{eq: P_eQ}
\end{equation}
When $M_\textrm{sc}$ is large enough, $\Tilde{P_e}$ approximates to $P_e$,
\begin{equation}
    \Tilde{P_e} \stackrel{\text{large } M_\textrm{sc}}{\longrightarrow} P_e.
\end{equation}
According to the LTE specification, the SRS has at least 24 subcarries\footnote{When LTE uplink bandwidth is 1.4 MHz, 6 Resource Block.}, which is rarely used in practice though. In most scenarios, the SRS occupies hundreds of subcarriers. For a rough BER performance estimation, approximating the Gaussian distribution is sufficient, as it simplifies the calculations and avoids the complexity of dealing with a series of regularized incomplete Beta functions.

\section{Coverage range analysis \label{sec:coverage}}
\begin{figure}
    \centering
    \subfloat[]{
    \includegraphics[width=0.75\columnwidth,trim={2cm, 0cm, 1.8cm, 0cm}, clip]{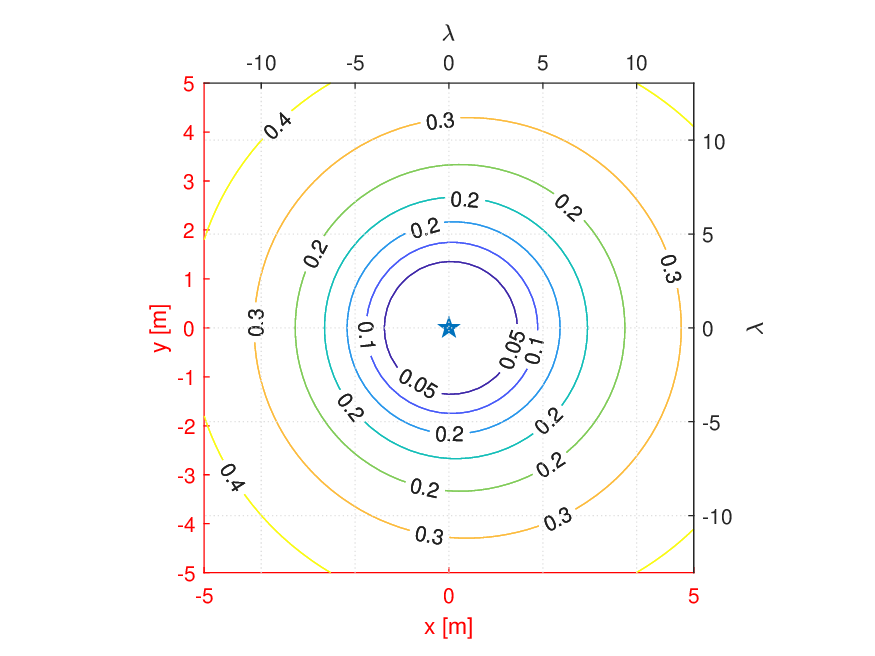}
    \label{fig:berContours}
    }
    \\
    \subfloat[]{
    \includegraphics[width=0.75\columnwidth,trim={2cm, 0cm, 1.8cm, 0cm}, clip]{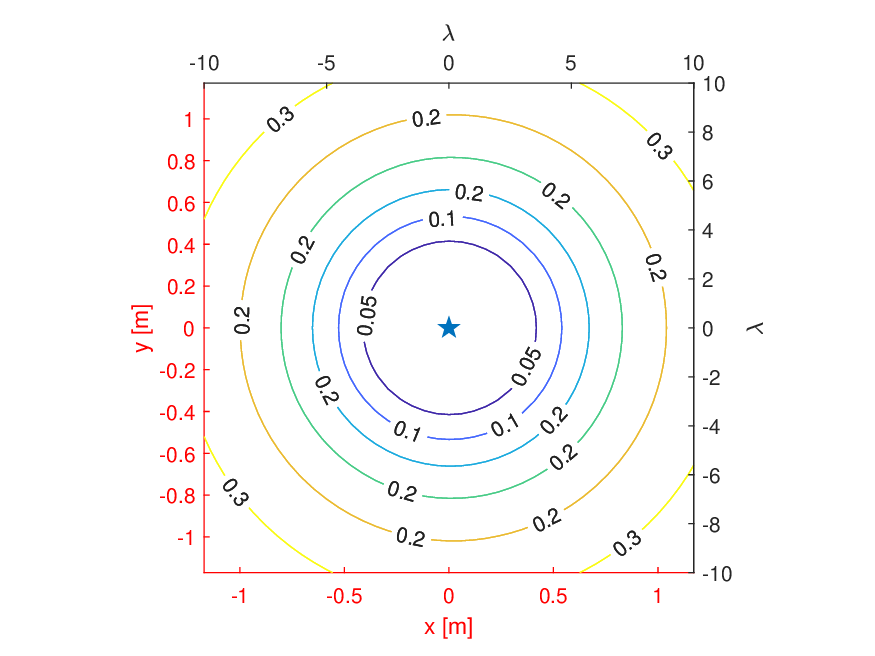}
    \label{fig:berContours2560MHz}
    }
    \\
    \subfloat[]{
    \includegraphics[width=\columnwidth,trim={0cm, 2.9cm, 1cm, 3cm}, clip]{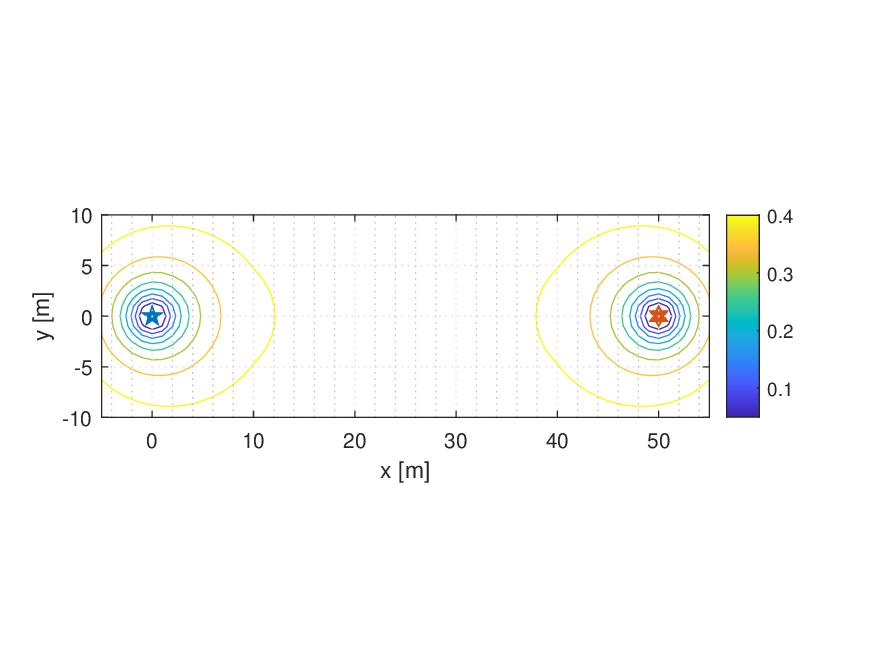}
    \label{fig:berContours_far}
    }
    \caption{Fix the UE at origin (0,0) blue pentagram and the BS at (50,0) orange hexagram, then move the BD around. The BER of AmBC shows as contours values. The received LTE SNR at BS is 10 dB. a) Coordinates nearby UE with 782 MHz LTE uplink; b) Coordinates nearby UE with 2560 MHz LTE uplink; c) Coordinates full range with 782 MHz LTE uplink.}
    \vspace{-15pt}
\end{figure}
As derived the BER in Eq. \eqref{eq: P_e}, AmBC performance is found to depend on the channel $h_d$, $h_s$, $h_b$, and the noise power $\sigma_n^2$. This naturally raises an interesting question: How far can the UE reliably receive the BD signal?

In practical scenarios, the LTE cellular environment remains relatively stable. This means that factors such as the environmental noise power $\sigma_n^2$, LTE central frequency, and the location of BS are independent of the backscatter communication. Assuming that the locations of the UE and the BS are fixed, while the BD moves around the UE, we investigate how the BER performance in relation to spatial coordinates.

The LTE SNR is supposed to $|S_t[k;l]|^2|h[l]|^2/\sigma_n^2$.
Because SRS symbol $|S_t[k;l]|^2$ always 1, LTE SNR is simplified as $|h[l]|^2/\sigma_n^2$. But in practice, the scattered path is much weaker than direct path, $|h_d|\gg|h_sh_b|$. So we roughly define LTE SNR as
\begin{equation}
    \gamma=\frac{|h_d|^2}{\sigma_n^2}.
    \label{eq: SNR_gamma}
\end{equation}
Once the distance between the UE and the BS is fixed, the SNR of the LTE link becomes constant.

The distance between the BS, UE and BD are denoted as $d_d$, $d_s$, $d_b$, respectively. Their subscripts correspond to similarly meanings as the channel $h_d$, $h_s$, $h_b$.
A simple Free-space Path Loss (FSPL) model is applied to calculate channel losses, which are expressed in terms of complex baseband signals, as outlined in \cite{saunders2007antennas}
\begin{equation}
    h_d = \frac{\lambda}{4 \pi d_d} \exp\left(j\frac{2 \pi d_d}{\lambda}\right),
\end{equation}
\begin{equation}
    h_s = \frac{\lambda}{4 \pi d_s} \exp\left(j\frac{2 \pi d_s}{\lambda}\right),
\end{equation}
\begin{equation}
    h_b = \frac{\lambda}{4 \pi d_b} \exp\left(j\frac{2 \pi d_b}{\lambda}\right),
\end{equation}
where $\lambda$ is wavelength of LTE cellular system. Additionally, we notice that BER performance depends on the ratio between the scatter path and direct path channel gains
\begin{equation}
    \iota=\frac{h_s h_b}{h_d} = \frac{\lambda}{4\pi}\frac{d_d}{d_sd_b}\exp\left[j\frac{2 \pi (d_d-d_b-d_s)}{\lambda}\right].
\end{equation}
When $\gamma$ is constant, the BER becomes a function related to $\left|1 + \iota\right|^2$
\begin{equation}
    \begin{split}
        P_e(\iota)= &\frac{1}{2} +
        \frac{1}{2}
        \exp \left[ -\frac{M_\mathrm{sc}N}{4}\gamma\left(1+\left|1 + \iota\right|^2\right) \right] +\\
    &\sum_{j=0}^\infty \sum_{k=0}^\infty
        \left(\frac{M_\mathrm{sc}N}{4} \gamma \right)^{j+k}
        \frac{\left|1 + \iota\right|^{2j} 
        -\left|1 + \iota\right|^{2k}}{j!k!}\\
        &\bar{B}_\frac{x}{1+x}
        \left[\frac{M_\mathrm{sc}N}{2}+j,\frac{M_\mathrm{sc}N}{2}+k\right],
    \end{split}
\label{eq: exactBERandIota}
\end{equation}
where
\begin{equation}
    \left| 1 + \iota \right|^2 = 1 + \frac{\lambda}{2\pi}\frac{d_d}{d_sd_b}
    \cos\left[2 \pi\frac{ d_d-d_b-d_s}{\lambda}\right]
    + \left(\frac{\lambda}{4\pi}\frac{d_d}{d_sd_b}\right)^2.
\end{equation}
The exact BER $P_e$ is too complex to fully understand the geometer shape of the BER contours from the analytical expression in Eq. \eqref{eq: exactBERandIota}. The Fig. \ref{fig:berContours}, \ref{fig:berContours2560MHz} and \ref{fig:berContours_far} show the $P_e$ for different BD locations based on numerical simulation. As shown in Fig. \ref{fig:berContours}, the BER contour near the UE is roughly circular.

In the reset of this section we demonstrate that the BER contour near the UE is roughly circular, both through analytic approximation and numerical simulation. The asymptotic BER $\Tilde{P_e}$ is
\begin{equation}
    \Tilde{P_e}(\iota) = Q\left(\sqrt{
    \frac{N M_\textrm{sc} (|1+\iota|^2-1)^2}{4(1+\gamma|1+\iota|^2+\gamma)}
    }\right).
\end{equation}
The relation between $\Tilde{P_e}$ and $\left| 1 + \iota \right|^2$ is
\begin{equation}
\begin{split}
    \left| 1 + \iota \right|^2 = &1 + \frac{2\gamma[Q^{-1}(\Tilde{P_e})]^2}{NM}\\
    &+\frac{2Q^{-1}(\Tilde{P_e})}{\sqrt{NM}} \sqrt{\frac{[\gamma Q^{-1}(\Tilde{P_e})]^2}{NM}+2\gamma+1},
\end{split}
\end{equation}
where $Q^{-1}(\cdot)$ refers to inverse Q-function.

Due to BD is nearby BS, $d_s\ll d_d \approx d_b$. The path loss between BD to BS has roughly the same amplitude of UE to BS. So their path loss model is not critical. The channel between UE and BD dominates $\iota$, approximating as
\begin{equation}
    \iota \approx \frac{\lambda}{4\pi d_s} \exp\left[-j\frac{2 \pi d_s}{\lambda}\right],
\end{equation}
and
\begin{equation}
    \left| 1 + \iota \right|^2 \approx 
    1 + \frac{\lambda}{2\pi d_s}
    \cos\left[-2 \pi\frac{d_s}{\lambda}\right]
    + \left(\frac{\lambda}{4\pi d_s}\right)^2.
\end{equation}
When BS is far away from the UE, $d_s$ becomes the sole independent variable for the $\Tilde{P_e}$ contour lines. In other words, the set of coordinates with the same $\Tilde{P_e}$ forms a circle, as the distance to the UE remains constant. 
Consequently, these $\Tilde{P_e}$ contour lines form concentric circles centered around the UE.

The bistatic and dislocated backscatter link budget was given by Griffin and Durgin \cite[Eq. (4)]{5162013}. The received modulated backscatter power depends on the distances between the BS, the UE and the BD, as discussed in this study. Additionally, it is influenced by factors such as antenna gains, arrival angles, and polarization mismatches. Further contributions to path loss include path blockage losses and small-scale fading losses. Specifically, the BD may exhibit mismatches between load impedance and antenna impedance, and transmission line impedance of circuits, leading to dissipation of scattered signal energy.

In this paper, the analysis is conducted under ideal conditions, with propagation loss assumed to dominate the link budget. The wavelength $\lambda$ is determined by the LTE cellular network. Isotropic antennas with unit gain and perfect polarization matching are assumed. The BD is modeled as an ideal implementation. Inevitably, the presence of the BD alters direct path channel, known as \textit{structural mode} of the BD. The resulting additional multipath components are incorporated in direct path channel, while those multipath components affected by the BD operation are treated as a part of scattered path. The effects of other parameters are neglected in this analysis.

The LTE is simulated operating at 782 MHz and 2560 MH with a 10 MHz bandwidth. This concentric-circle behaviors of the BER is valid only when the BS is significantly distant from the UE and the BD is relatively close to UE. In Fig. \ref{fig:berContours} an Fig. \ref{fig:berContours2560MHz}, the BER contours in the vicinity of the UE are nearly circular, though their centers are slightly shifted towards the BS. In Fig. \ref{fig:berContours_far}, BER contours deviate form circular shapes when the BD is relatively far from the UE. The simulations in Fig. \ref{fig:berContours2560MHz} and Fig. \ref{fig:berContours} show a proportional relationship between $P_e$ and $\lambda$. In network planning, a BER threshold of less than $10^{-1}$ is considered; otherwise, the communication is deemed infeasible and ineffective. The detection range is proportional to the wavelength. Under this constrain, the BER contour lines can be approximated as concentric circles, indicating a reliable communication range of approximately four wavelength between the UE and the BD. Based on both analytic expression and numerical simulation, the BER circular approximates to circle for low BER cases around UE.

\section{Simulation and measurement \label{sec:Simulation}}
In pervious section, we approximated the optimal receiver described in Eq. \eqref{eq:BesselMAP} using various approaches. In this section, we validate different receivers BER performances through numerical simulations, and further demonstrate the feasibility of AmBC based on LTE uplink via over-the-air measurements.

A common LTE setup and parameters are employed for both the numerical simulations and the over-the-air measurements.
The uplink utilizes SC-FDMA with 50 Resource Block (RB), 15 kHz subcarrier spacing and a normal Cyclic Prefix (CP). The SRS are transmitted periodically every 2 ms, occupying the entire bandwidth of the last OFDM symbol in a subframe. In other words, SRS provides 500 Hz sampling rate for AmBC. The number of SRS subcarrier, $M_\textbf{sc}$, is constant at 288 in this configuration.
The UE, BS and BD have only single antenna.

\subsection{Numerical simulation}
In all numerical simulations is based on following parameters. The LTE system operates on a 782 MHz uplink frequency with a 10 MHz bandwidth. The direct path attenuation is -52.2 dB and scattered path attenuation is -82.6 dB. The symbol length of BD signal is 8 ms, that is $N=4$. BPSK symbols are $\boldsymbol{s_0}=[-1,1,-1,1]$ and $\boldsymbol{s_1}=[1,-1,1,-1]$. FSK symbols frequencies are 125 Hz and 250 Hz, $\boldsymbol{s_0}=[-1,-1,1,1]$ and $\boldsymbol{s_1}=[-1,1,-1,1]$ respectively.

In this section, we refer to hypothesis test in Eq. \eqref{eq: rxCorr} as correlation receiver, hypothesis test in Eq. \eqref{eq: rxSqrt} as square root receiver and hypothesis test in Eq. \eqref{eq: square} as square receiver. 

In this subsection, we firstly give the simulation assumptions and parameters. And we evaluate the whole AmBC with respect to BER performance using BPSK, FSK and D-BPSK
The numerical simulation discussed to support propositions:
\begin{enumerate}
    \item The theoretical matches simulated BER performance of AmBC.
    \item Three receivers have equivalent BER performance.
    \item Gaussian is a good approximation of noncentral chi-square distributed $y$ and doubly noncentral F distributed $\xi$.
\end{enumerate}

\subsubsection{Overall system BER simulation}
\begin{figure}
    \centering  
    \includegraphics[width=\columnwidth]{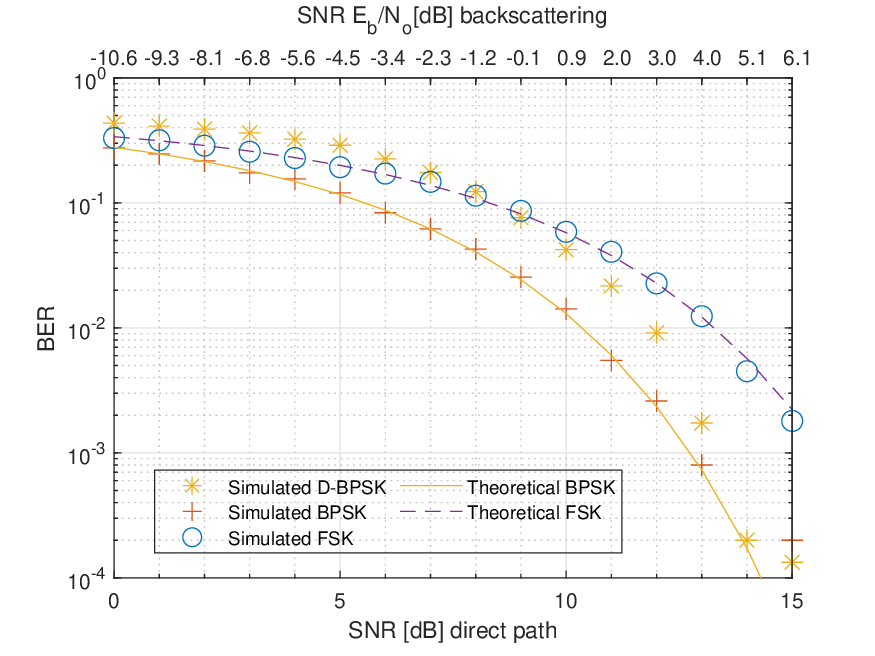}
    \caption{AmBC BER simulations with overall LTE system. The figures x-axis are SNR in LTE cellular network and SNR in AmBC.}
    \label{fig:wholeSystem}
    \vspace{-15pt}
\end{figure}

The BER performance of AmBC based on LTE cellular network are simulated using resource grid with AWGN, as shown in Fig. \ref{fig:wholeSystem}. For simulated points, the BD transmitted 10,000 symbols with equal symbol probabilities to calculate the average BER. The theorem BER curves are correlation receivers as shown Eq. \eqref{eq: P_e}, which fit the Monte Carlo experiments. The lower x-axis represents the SNR, $\gamma$, as defined in Eq. \eqref{eq: SNR_gamma}. And the upper x-axis represents the SNR per bit, $\gamma_b$, as defined in Eq. \eqref{eq: SNR_gamma_b}. It is evident that BPSK offers superior BER performance compared to other modulation schemes. However, slow channel fading causes sign flipping of $\Delta h$, erupting a burst of BPSK errors due to its detection reliance on strict phase coherence. To mitigate this, FSK and Differential BPSK (D-BPSK) modulation schemes are employed. While D-BPSK performs worse than FSK at lower SNR, the it outperforms FSK when target BER is below $10^{-1}$, which is a reasonable threshold for reliable communication. Unfortunately, the D-BPSK theoretical BER is not provided because cross-correlation functions of noncentral chi-square R.V.s are not clear and needs further study.


\subsubsection{Receiver performance comparison}
\begin{figure}
    \centering  
    \includegraphics[width=\columnwidth]{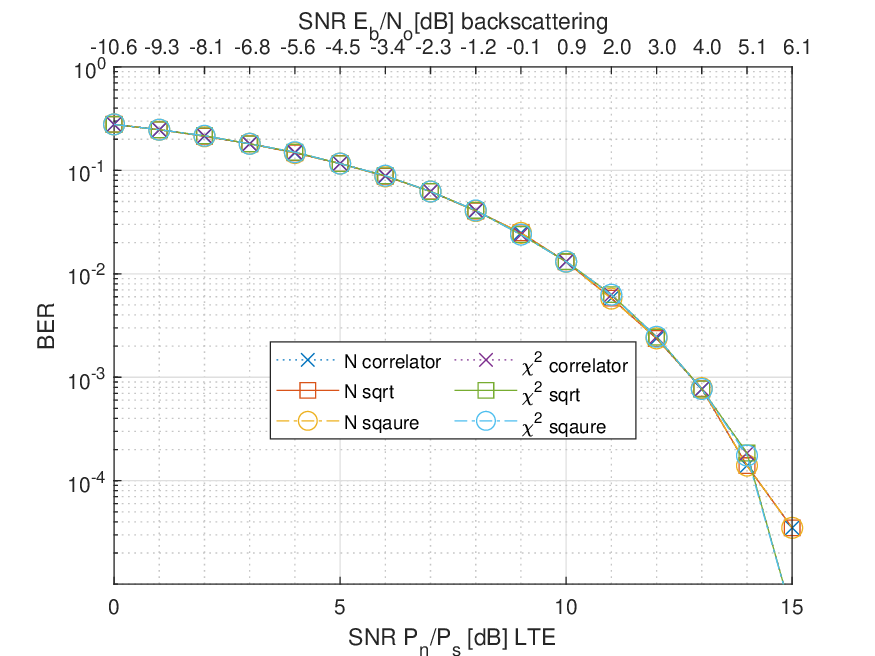}
    \caption{$y$ is noncentral chi-square distribution (labeled '$\chi^2$') and asymptotes to normal distribution due to CLT (labeled 'N'). Generate noncentrial chi-square R.V. $y$ and is detected by different receiver, in Monte Carlo method, BPSK symbol. The label correlation is Eq. \eqref{eq: rxCorr}; label sqrt is Eq. \eqref{eq: rxSqrt}; label square is Eq. \eqref{eq: square}.}
    \label{fig:yChi&N}
    \vspace{-15pt}
\end{figure}

This simulation compared the BER performance of various receivers using the BPSK modulation scheme. The analysis generates the noncentral chi-square with $2M_\textrm{sc}$ degrees of freedom R.V. $y$ in Eq. \eqref{eq:yChiSquare} with 100,000 realizations, labeled '$\chi^2$' in the Fig. \ref{fig:yChi&N}. The optimal receiver is defined by Eq. \eqref{eq:BesselMAP}, however its reliance on high-order Bessel functions poses significant challenges for numerical computation. To address this, the optimal receiver is approximated using three simpler receivers: square root receiver (Eq. \eqref{eq:SqrtMAP}), the correlation receiver (Eq. \eqref{eq:corrMAP}) and the square receiver (Eq. \eqref{eq:GaussianMAP}). The Fig. \ref{fig:yChi&N} compares the BER performance of these three approximations. The results indicates that all three approximations exhibit nearly equivalent BER performance. Among them, the correlation receiver is the most simplest to implement in hardware for demodulation purposes.

Then the simulation also demonstrates that the Gaussian asymptotes accurately noncentral chi-square, with comparable performance. The $y$ is modeled as a Gaussian distribution in Eq. \eqref{eq:yGaussian}, labeled `N' in the Fig. \ref{fig:yChi&N}. For Gaussian $y$, the square receiver serves as the optimal receiver. As shown in the Fig. \ref{fig:yChi&N}, the results intuitively confirm that the square root receiver and the correlator receiver perform as well as optimal receiver under Gaussian conditions. This validates the reliability of approximating of the square root and correlation receiver to the square receiver in the Gaussian case. Furthermore, when comparing with Gaussian $y$ and noncentral chi-square $y$, the BER performance of chi-square with $2M_\textrm{sc}$ degrees of freedom asymptotes closely to that of the Gaussian distribution. Thus, we conclude that Gaussian approximation is sufficiently accurate to replace chi-square with with $2M_\textrm{sc}$ degrees of freedom in this context.

\subsubsection{Theorem BER}
\begin{figure}
    \centering  
    \includegraphics[width=\columnwidth]{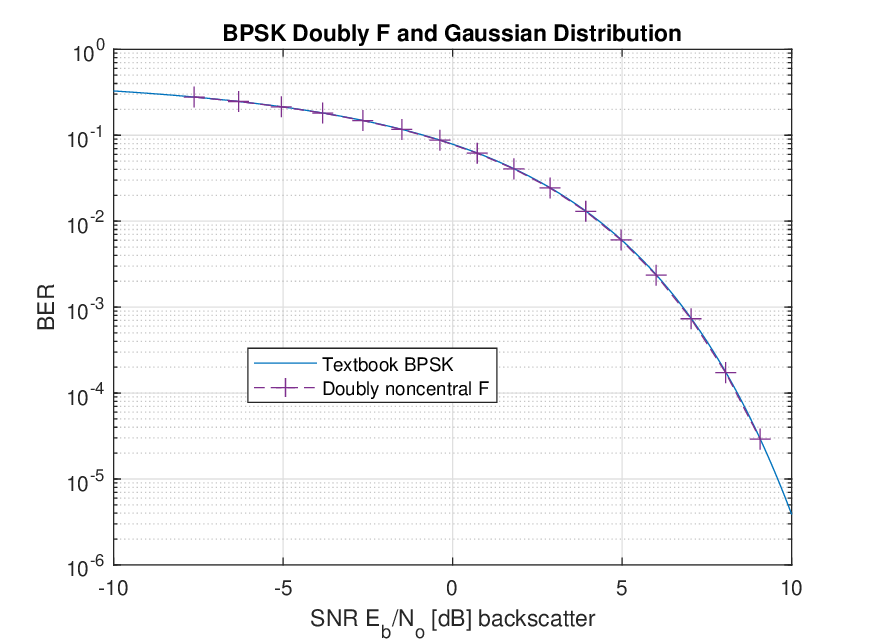}
    \caption{The doubly noncentral F Eq. \eqref{eq: P_e} BER approximates to Gaussian Eq. \eqref{eq: P_eQ} BER (Textbook BPSK), when $M_\textrm{sc}$ is large enough.}
    \vspace{-10pt}
    \label{fig:doublyF_to_Qfunc}
\end{figure}

In Fig. \ref{fig:doublyF_to_Qfunc}, the doubly noncentral F error possibility $P_e$ is compared with asymptotic Gaussian error possibility $\Tilde{P_e}$. The results demonstrate that, in LTE scenario, the doubly noncentral F distribution close approximates the Gaussian distribution. Additionally, the bell-shaped \textit{p.d.f.} of doubly noncentral F distribution, as shown in Fig. \ref{fig:pdfDoublyNoncentralFDistribution}, offers another perspective for drawing a connection to the Gaussian \textit{p.d.f}.

\subsection{Over-the-air Measurement}
\begin{figure}
    \centering
    \subfloat[]{
    \includegraphics[width=\columnwidth]{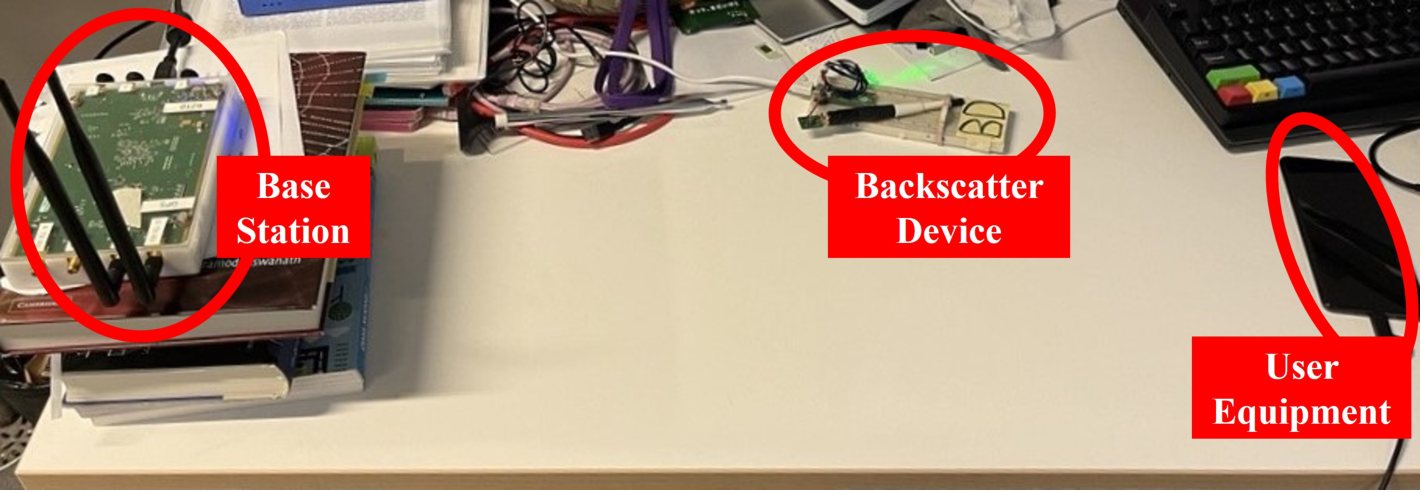}
    \label{fig:measurementSetup}
    }
    \\
    \subfloat[]{
    \includegraphics[width=\columnwidth]{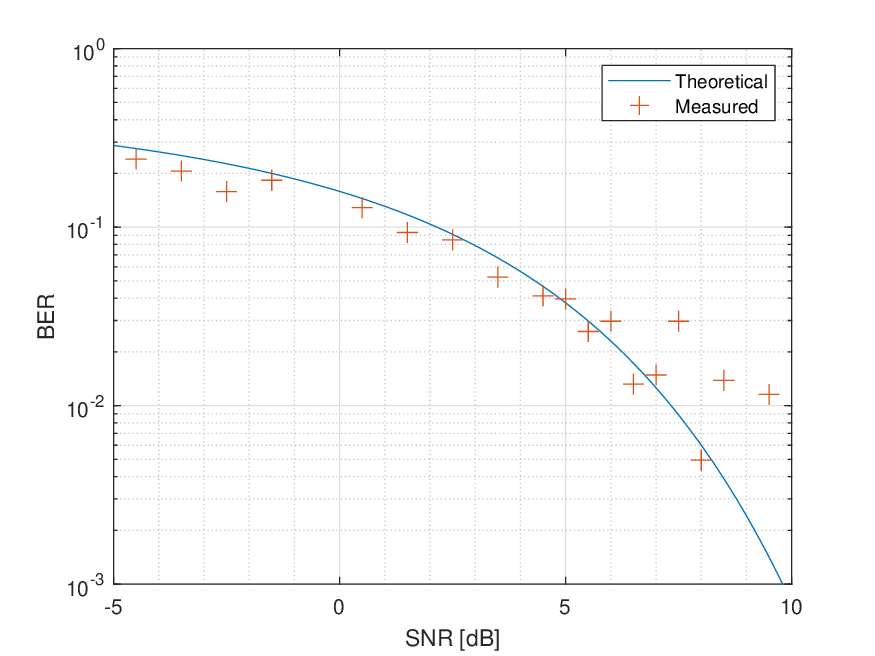}
    \label{fig:measurement}
    }
    \caption{ Over-the-air measurement of FSK AmBC using coherent detector. a) illustration of measurement setup; b) measured BER performance.}
    \vspace{-15pt}
\end{figure}

We setup a bistatic AmBC system based on LTE uplink, as illustated in Fig. \ref{fig:measurementSetup}. The measurement campaign was conducted in an indoor environment at Aalto University TUAS building. The distance between those devices are $d_d = 0.83 \textrm{ m}$, $d_s = 0.45 \textrm{ m}$ and $d_b = 0.65 \textrm{ m}$. The LTE uplink operates at a central frequency of 2560 MHz with a bandwidth of 10 MHz, as 782 MHz is not supported by our BS. However, since the over-the-air measurements focus on the relationship between BER and LTE SNR, the central frequency does not impact the validity of our experiment. As introduced at the beginning of this section, the SRS is transmitted in every 2 ms by a commercial UE, specifically the Samsung SM-G9860.

The BS is implemented using a Software-Defined Radio (SDR) based on the srsRAN Project LTE BS framework \cite{gomez2016srslte} and the Cumucore core network \cite{s21113773}. A National Instrument USRP B210 serves as the RF front end, equipped with 12-bit A/D and D/A converter, and a sampling rate of 30.72 MHz. The received signal is adjusted appropriately to fit within the dynamic range of the USRP.

The BD implementations modulation with a return loss of 0.5 dB for reflection and 23 dB for absorption, as described in \cite{9464748}. It is controlled by an RP2040 microcontroller chip, which generates a square wave. The BD toggles between two frequencies, 250 Hz and 125 Hz, to create an artificial Doppler higher than natural Doppler frequency. The modulation employs a 40 ms symbol period using square-wave FSK. A single BD frame consists of a 21-bit synchronization header followed by 80 information bits. To synchronize the start sample, three 7-bit Barker codes are used. Frames are transmitted continuously in a loop, with each 4.04 s frame followed by 0.5 s idle period.

The AmBC receiver, implemented in Matlab, utilizes the uplink channel estimation from BS. A coherent FSK detector demodulates the BD samples based on LTE channel estimates. The theoretical BER for a coherent FSK detector is given by textbook \cite{proakis2002communication} Eq. (5.2-11)
\begin{equation}
    P_{e\textrm{FSK}} = Q\left(\sqrt{\frac{\mathcal{E}_b}{N_0}}\right),
\end{equation}
and is labeled as `theoretical' in Fig. \ref{fig:measurement}. The complete over-the-air measurement setup is depicted in Fig. \ref{fig:measurementSetup}.

The Fig. \ref{fig:measurement} illustrates the over-the-air measurements, with the theoretical BER curve included as a reference. The measured data were post-processed to account for the discrete nature of BER calculations, because each BD packet only 101 bits. And packets without any error were not represented in BER plot. To align the theoretical BER, the results were smoothed by averaging errors across packets with specific SNR intervals. The SNR of each packet was recorded, and packets were grouped into 0.25 dB SNR intervals. The bit errors within each 0.25 dB SNR interval were then aggregated to compute a single BER value for that interval. The smoothed BER points are shown in Fig. \ref{fig:measurement}.

As shown in Fig. \ref{fig:measurement}, the measured data points align closely with the theoretical BER curve, with most points concentrated in the SNR range of 5 dB to 8 dB. This result demonstrates the feasibility of AmBC based on LTE uplink. The long square-wave FSK symbol enables the BD receiver to accumulate sufficient energy from the scattered path to reliable detect the BD signal. Practical over-the-air experiments revealed that a 40 ms symbol period achieves a reasonable trade-off between AmBC BER performance and FSK symbol duration. The measured BER is on the order of magnitude $10^{-2}$, which is adequate for many AmBC applications.

\section{Conclusion \label{sec:Conclusion}}
In this paper, we discuss the opportunities of LTE uplink in the aspect of AmBC. We compare accurate and approximate BER performance. Then we analyze the AmBC coverage. The numerical results shows that Gaussian approximation is good enough to analyze BER in context of AmBC. We conduct a proof-of-concept experiment to show the feasibility of AmBC in LTE uplink. The theoretical BER analyze is corroborated by over-the-air measurement results.
Combined with our LTE downlink AmBC study \cite{10136397}, the AmBC works as a symbiotic radio in both downlink and uplink in LTE cellular system. 

\section{Acknowledgments}
This work is in part supported by the European Project Hexa-X II under (grant 101095759) and Business Finland Project eMTC (Dnro 8028/31/2022).

\bibliographystyle{IEEEtran}
\bibliography{References}

\end{document}